\documentclass[twocolumn]{aa}  

\DeclareUnicodeCharacter{202F}{\,}
\usepackage{graphicx}
\usepackage{lipsum}
\usepackage{subcaption}         
\usepackage{lscape}             
\usepackage{placeins}           

\usepackage{amsmath}
\usepackage{amssymb}
\usepackage{lipsum}
\usepackage{graphics}
\usepackage[varg]{txfonts}
\usepackage{graphicx}
\usepackage{epsfig}
\usepackage{wrapfig}
\usepackage{rotating}
\usepackage{color}
\usepackage{comment}
\usepackage[normalem]{ulem}
\usepackage{multirow}
\usepackage{tabularx}
\usepackage{graphicx}
\usepackage{dcolumn}
\usepackage{hyperref}
\hypersetup{colorlinks=true, allcolors=blue}
\usepackage{float}
\usepackage{natbib}

\usepackage{pgfplots}
\usetikzlibrary{pgfplots.groupplots}
\usepackage{tikz}
\usetikzlibrary{positioning}
\usetikzlibrary{calc}
\usepackage{fp}
\usetikzlibrary{fixedpointarithmetic}

\usepackage{siunitx}
\usepackage{tablefootnote}
\usepackage{threeparttable}
\newlength\figureheight 
\newlength\figurewidth 
\usepackage{color}
                              
\begin{document}



%

\title{MAXI J1820+070: A rapidly spinning black hole with a small disk truncation in the soft state and warm corona layers}

\author{Th. V. Papavasileiou\inst{1, 2}\fnmsep\thanks{Corresponding author: th.papavasileiou@uowm.gr}
\and T. S. Kosmas\inst{1}
\and O. Kosmas\inst{3} 
\and I. Sinatkas\inst{2}}
\institute{Department of Physics, University of Ioannina, GR-45110 Ioannina, Greece \\
\email{hkosmas@uoi.gr}
\and Department of Informatics, University of Western Macedonia, GR-52100 Kastoria, Greece \\
\email{th.papavasileiou@uowm.gr, isinatkas@uowm.gr}
\and Trumeter Ltd, Pilot Mill, Alfred St, Bury,  BL9 9JR Bury, U.K. \\
\email odykosm@gmail.com}

\date{Received <date> /
Accepted <date>}


\abstract
{{Context. MAXI J1820+070 is a black hole X-ray binary discovered in 2018 during a major outburst that triggered extensive multiwavelength observations. Subsequent studies have focused on its physical properties, particularly the black hole spin. In this context, continuum fitting of the source spectra below 25 keV favors low-spin values, whereas spin estimates based on the relativistic precession model and reflection spectroscopy suggest a rapidly rotating black hole.}        
 
{Aims. Our study seeks to address the debate over the spin of MAXI J1820+070 through broadband spectral modeling of NICER/NuSTAR observations obtained during the soft state. We further compare our results with previous spin estimates and examine the source variability across the soft state. Finally, we investigate the origin of the soft X-ray excess, which we argue does not originate from the plunging region as previously suggested.}        

{Methods. We perform joint spectral fitting of NICER and NuSTAR data over the broadest possible energy range of $0.2\text{--}79\text{ keV}$. To achieve high accuracy in our spin estimation, we adopt a physically motivated model combination. We employ \texttt{kerrbb2} to describe the relativistic thin accretion disk, utilizing pre-calculated \texttt{bhspec} tables for the color-correction factor, convolved with \texttt{thcomp} to model the high-energy Comptonized tail. In addition, we modify our model to account for the substantial excess emission reported below 10 keV and conduct an MCMC analysis, thereby improving the statistical robustness of our fits. To further investigate the origin of the excess emission, we calculate spin-dependent radial disk temperature profiles across all epochs.} 

{Results. In this work, we show that the soft spectral state of MAXI J1820+070 is well-described by the Comptonized emission from a thin relativistic accretion disk and two blackbody spectral components with $kT_{\text{bb}} \approx 0.5\text{ keV}$ and $kT_{\text{bb}} \approx 0.2\text{ keV}$, respectively. Our results indicate that the black hole in MAXI J1820+070 is rapidly spinning ($a > 0.8$), potentially powering the observed relativistic jets. Our analysis reveals a significant decline in the inner disk temperature midway through the soft state, accompanied by a modest increase in the inferred inner disk radius up to $3R_{\text{g}}$. This behavior is consistent with a small disk truncation, a phenomenon recently predicted by GRMHD simulations of the plunging region. The two blackbody components are considerably cooler than the inner disk, corresponding to large emission regions of $\Delta R \approx 6R_{\text{g}}$ ($75\text{ km}$) and $\Delta R \approx 35R_{\text{g}}$ ($440\text{ km}$). This suggests that this emission may originate from two warm corona layers located beyond $\sim 10R_{\text{g}}$ and $\sim 40R_{\text{g}}$, in close analogy to the warm Comptonization models proposed to explain the soft X-ray excess in active galactic nuclei.}}



\titlerunning{MAXI J1820+070: A rapidly spinning black hole with a small disk truncation in the soft state and warm corona layers} 

\maketitle

\nolinenumbers

\section{Introduction}

MAXI J1820+070 is a black hole X-ray binary (BHXB) initially discovered in March 2018 as the optical transient ASASSN-18e \citep{Tucker_2018}. It was subsequently identified as the optical counterpart of an X-ray transient detected by the MAXI all-sky X-ray monitor aboard the ISS \citep{Kawamuro_2018, Denisenko_2018}. The source was later confirmed as a BHXB \citep{Uttley_2018, Baglio_2018a} and associated with a highly variable relativistic jet via optical and infrared variability analysis \citep{Bright_2018, Baglio_2018b, Bright_2020}.

The major outburst in 2018, leading to the source's discovery, remains the primary event providing a wealth of observational data over a timespan of several months. During this period, the source maintained its brightness while undergoing a full spectral state transition \citep{Shidatsu_2018, Shidatsu_2019, Russell_2019a}. The outburst was so bright, momentarily making MAXI J1820+070 the second brightest X-ray source in the sky. Several smaller flaring episodes followed in 2019 and 2020, leading to weaker activity reported in the following years \citep{Baglio_2019, Bahramian_2019, Ulowetz_2019, Williams_2019, Hambsch_2019, Adachi_2020, Sasaki_2020, Homan_2023, Baglio_2023}.

MAXI J1820+070 is located at a distance of $d=2.96\pm 0.33\text{ kpc}$ and has an inclination of $i=63\pm 3^{\circ}$, based on radio parallax measurements \citep{Atri_2020}. Adopting those values, optical spectroscopic analyses yield the black hole and the stellar companion masses as 8.48$^{+0.79}_{-0.72}$\(M_\odot\) and 0.61$^{+0.13}_{-0.12}$\(M_\odot\), respectively \citep{Torres_2020}.

The spin of the black hole remains a subject of debate among researchers. Reflection spectroscopy of NICER data obtained during the hard state favors low to moderate spin values \citep{Buisson_2019}, a result also supported by \citet{Atri_2020}.
Spectral analysis of NICER and Insight-HXMT data in the $1\text{--}25\text{ keV}$ band similarly suggests a low-spin black hole with $a < 0.3$. Specifically, \citet{Zhao_2021}, fitting Insight-HXMT observations in the $2\text{--}25\text{ keV}$ range, derived a spin of $a = 0.14\pm 0.09$, while continuum-fitting of a similar dataset by \citet{Guan_2021} yielded $a = 0.2^{+0.2}_{-0.3}$. More recent NICER analyses in the $1\text{--}10\text{ keV}$ band indicate $a = 0.38\pm 0.12$, whereas joint fitting with Insight-HXMT data in the $1\text{--}25\text{ keV}$ range suggests a slightly lower value of $a = 0.2\pm 0.1$ \citep{Sai_2024}.

Nevertheless, the continuum-fitting results appear to be contradicted by other spin-estimation methods \citep{Rarras_2024}. Timing analysis of the NICER data using the relativistic precession model yields a significantly higher black hole spin of $a = 0.799^{+0.016}_{-0.015}$ \citep{Bhargava_2021}. Furthermore, multi-epoch NuSTAR analyses employing relativistic reflection spectroscopy indicate that MAXI J1820+070 hosts a near-maximally rotating black hole with $a = 0.988^{+0.006}_{-0.028}$ and a disk inclination of $i=64^{+3}_{-4}$, consistent with previous estimates \citep{Draghis_2023}.      

A key distinction among these studies lies in their instrumental datasets and the specific energy ranges covered. Continuum-fitting spin estimates, relying solely on NICER and Insight-HXMT observations, focus below 25 keV, omitting the high-energy Comptonized emission that reveals how the disk dynamically interacts with the corona \citep{Papavasileiou2021, Papavasileiou_2023a, Papavasileiou_2023b, Kosmas_2023}. Conversely, standalone NuSTAR spectral analyses lack coverage below 3 keV, thereby missing the critical low-energy band where the majority of the disk continuum resides.  

Furthermore, the majority of the continuum-fitting estimations are derived on the basis of a relativistic thin accretion disk, whose emission is partially scattered in a power-law tail above 10 keV \citep{Steiner_2009}. However, this model does not account for the significant excess emission reported below 10 keV \citep{Fabian_2020}, which can lead to increased parameter uncertainties and inflated residuals. \citet{Fabian_2020} modeled the soft X-ray as a blackbody spectrum with $kT = 1\text{ keV}$, which is hotter than the inner disk assuming a low spin of $a = 0.2$. Therefore, the emission was speculated to originate near the edge of the plunging region due to magnetic stresses at the innermost stable circular orbit (ISCO) \citep{Mummery_2024}.

In this work, we aim to shed light on the spin controversy of MAXI J1820+070 by fitting the entire $0.2\text{--}79\text{ keV}$ NICER/NuSTAR bandpass during the 2018 outburst. We perform statistically robust analysis by properly modeling the excess emission below 10 keV and examining the source evolution throughout the soft state, including the potential origin of the blackbody-like emission. To that end, we calculate the radial temperature profile of the accretion disk and its evolution throughout the soft state, based on the spin-dependent model introduced in our previous work \citep{Papavasileiou_2025}. 

In the following section (Sec. \ref{Sec2}), we describe the data reduction and spectral extraction process. Sec \ref{Sec3} presents our best-fit results and spectral analysis, while the subsequent sections provide a detailed discussion of the implications of our findings. Finally, we summarise our conclusions in Sec. \ref{Sec7}.
    


\section{Observational data reduction}\label{Sec2}

\subsection{NuSTAR}

For our analysis, we selected the NuSTAR observations obtained in July and August 2018, when MAXI J1820+070 was going through the soft state. The observation IDs and their respective exposure times are listed in Table \ref{Table1}. This table also presents the epoch abbreviations adopted throughout this work.

\begin{table}
\caption{\label{Table1} Observational data from NuSTAR and NICER during the soft state of MAXI J1820+070. The exposure time was averaged over both FPMA and FPMB modules.}
\begin{center}
\begin{tabular}{l l l l l}
\hline \\ [0.005ex] 
OBSID & Abbreviation & Exposure time & Count rate \\ 
\hline \\ [0.01ex]
\multicolumn{4}{c}{NuSTAR} \\ [0.2ex]
\hline \\ [0.01ex]
90401309023 & nu23 & 7.307 \, ks & 750 s$^{-1}$ \\ [0.2ex]
90401309025 & nu25 & 5.420 \, ks & 570 s$^{-1}$ \\ [0.2ex]
90401309027 & nu27 & 17.509 ks & 437 s$^{-1}$ \\ [0.2ex]
90401309029 & nu29 & 6.938 \, ks & 303 s$^{-1}$ \\ [0.2ex]
90401309031 & nu31 & 17.045 ks & 255 s$^{-1}$ \\ [0.2ex]
\hline \\ [0.01ex]
\multicolumn{4}{c}{NICER} \\ [0.2ex]
\hline \\ [0.01ex]
1200120207 & ni207 & 4.596 ks & 25392 s$^{-1}$ \\ [0.2ex]
1200120220 & ni220 & 1.738 ks & 22678 s$^{-1}$ \\ [0.2ex]
1200120236 & ni236 & 1.825 ks & 18672 s$^{-1}$ \\ [0.2ex]
1200120240 & ni240 & 4.096 ks & 17184 s$^{-1}$ \\ [0.2ex]
\hline
\end{tabular}
\end{center}
\end{table} 

We reduced the data obtained from the \texttt{HEASARC} archive using the most recent \texttt{caldb} calibration files and the \texttt{nustardas} pipeline tool via the \texttt{nupipeline} command. During processing, we applied strict filtering to exclude intervals when the spacecraft passed through the South Atlantic Anomaly (SAA), enabling the \texttt{saamode} option and activating the \texttt{tentacle} flag calculation. Additionally, we applied the standard strict screening criteria, including the expression \texttt{(STATUS==b0000xxx00xxxx000)\&\&(SHIELD==0)}, to guarantee high data quality. The recommended \texttt{SHIELD==0} filter was included to remove cosmic-ray–contaminated events.

\begin{figure}[ht] 
\begin{tikzpicture}
\centering
  \node (img1)  {\includegraphics[width=\linewidth]{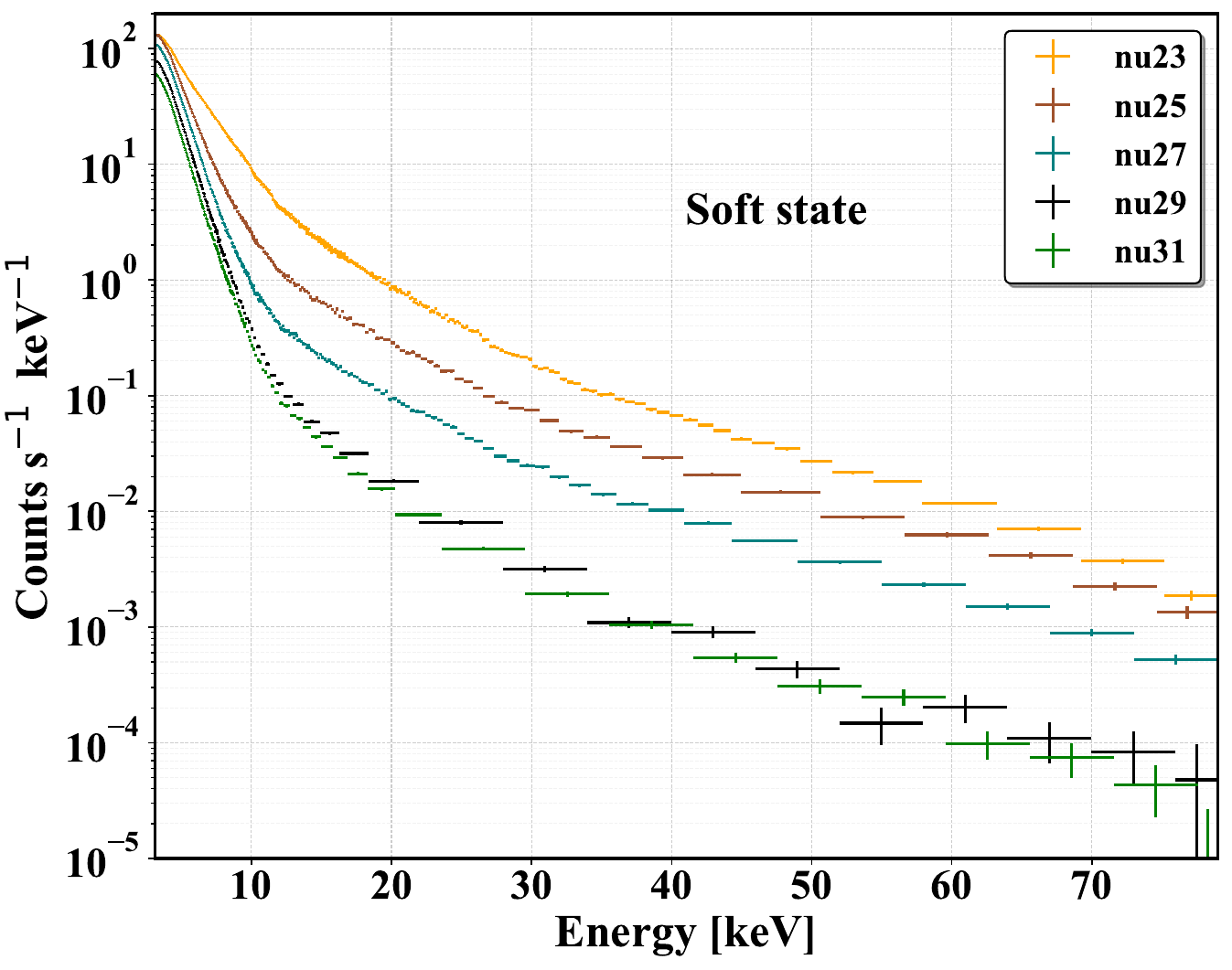}};
\end{tikzpicture}

\caption{\label{figure1} The merged spectra recorded by the two NuSTAR modules from MAXI J1820+070 in the soft state during the 2018 outburst.}
\end{figure}

After generating the clean event files for both NuSTAR modules (FPMA and FPMB), we used the \texttt{nuproducts} command to extract the source and background spectra. The source spectra were obtained from a circular region centered on the source coordinates with a radius of 50 arcsec, while the background spectra were extracted from an annular region with inner and outer radii of 123 and 197 arcsec, respectively. Nonetheless, the background contribution was negligible compared to the source brightness which dominated the spectra across the entire bandpass. 

The merged FPMA and FPMB spectra for all epochs are presented in Fig. \ref{figure1}. To secure optimal quality, the data were grouped using the optimal binning method of \citet{Kaastra_2016} for all spectral fitting. For visualization purposes, the spectra were further rebinned while plotting Fig. \ref{figure1}. Notably, epochs nu29 and nu31 exhibit many spectral similarities since the observations occurred only a few days apart.

\subsection{NICER}

To extend the energy coverage of our analysis, we include the NICER data obtained simultaneously or quasi-simultaneously with the NuSTAR observations. These correspond to OBSIDs 1200120207, 1200120220, and 1200120240 (listed in Table \ref{Table1}), providing continuous $0.2\text{--}79\text{ keV}$ broadband spectral coverage for NuSTAR epochs nu25, nu27, and nu31, respectively. In addition, we include an observation (ni236) obtained during a preceding time interval relative to the nu29 epoch as a validation dataset.

Level 2 and 3 spectral products were reduced using the \texttt{nicerl2} and \texttt{nicerl3-spect} tools within the \texttt{nicerdas} software package. Notably, our data are entirely unaffected by the well-documented optical light leak impacting daytime NICER observations, as these sessions occurred prior to the May 2023 shield breach. Furthermore, we utilized the latest \texttt{HEASOFT} version and calibration files to ensure adherence to all current data-filtering recommendations from the NICER team.

\section{Best-fit results and spectral analysis}\label{Sec3}

We consider the entire $0.2\text{--}79\text{ keV}$ broadband energy range in our joint fits. Specifically, we fit epoch nu25 simultaneously with ni207, nu27 with ni220, and nu31 with ni240. For these joint sets, we utilize the $0.2\text{--}5\text{ keV}$ NICER bandpass and the $5\text{--}79\text{ keV}$ NuSTAR range due to the well-documented calibration discrepancy between the two instruments, wherein NICER exhibits a steeper flux drop above 5 keV. All model parameters are tied between the respective NICER and NuSTAR spectra, leaving only a cross-calibration multiplication constant free to vary. The NICER background is modeled using the SCORPEON framework. The remaining NuSTAR-only observations are fitted independently across their full $3\text{--}79\text{ keV}$ bandpass. For every individual epoch, we jointly fit the spectra from both NuSTAR focal plane modules, allowing a cross-calibration constant to float between FPMA and FPMB. 

\begin{figure}[ht] 
\begin{tikzpicture}
\centering
  \node (img1)  {\includegraphics[width=\linewidth]{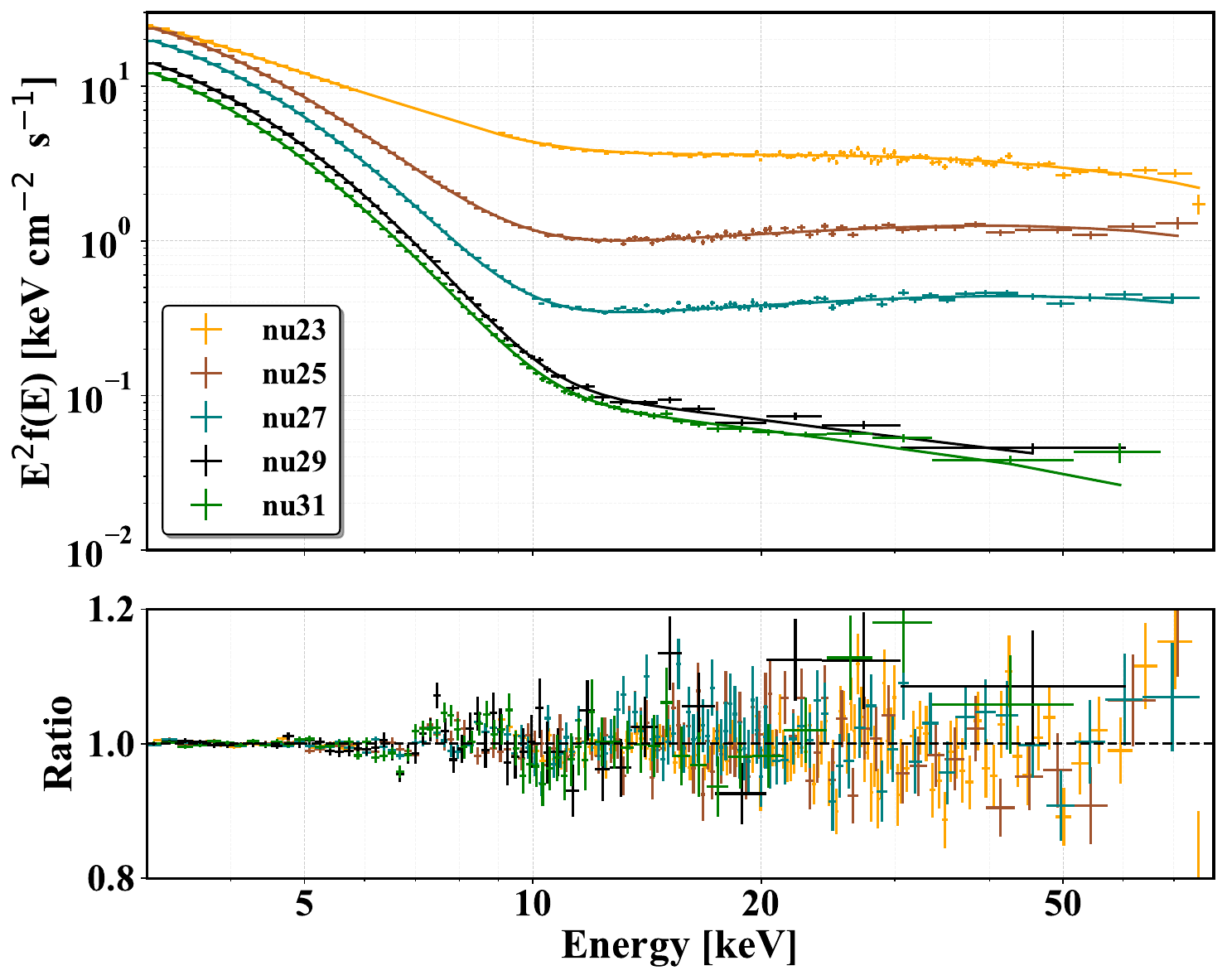}};
  \node[below= of img1, yshift=1cm] (img2)  {\includegraphics[width=\linewidth]{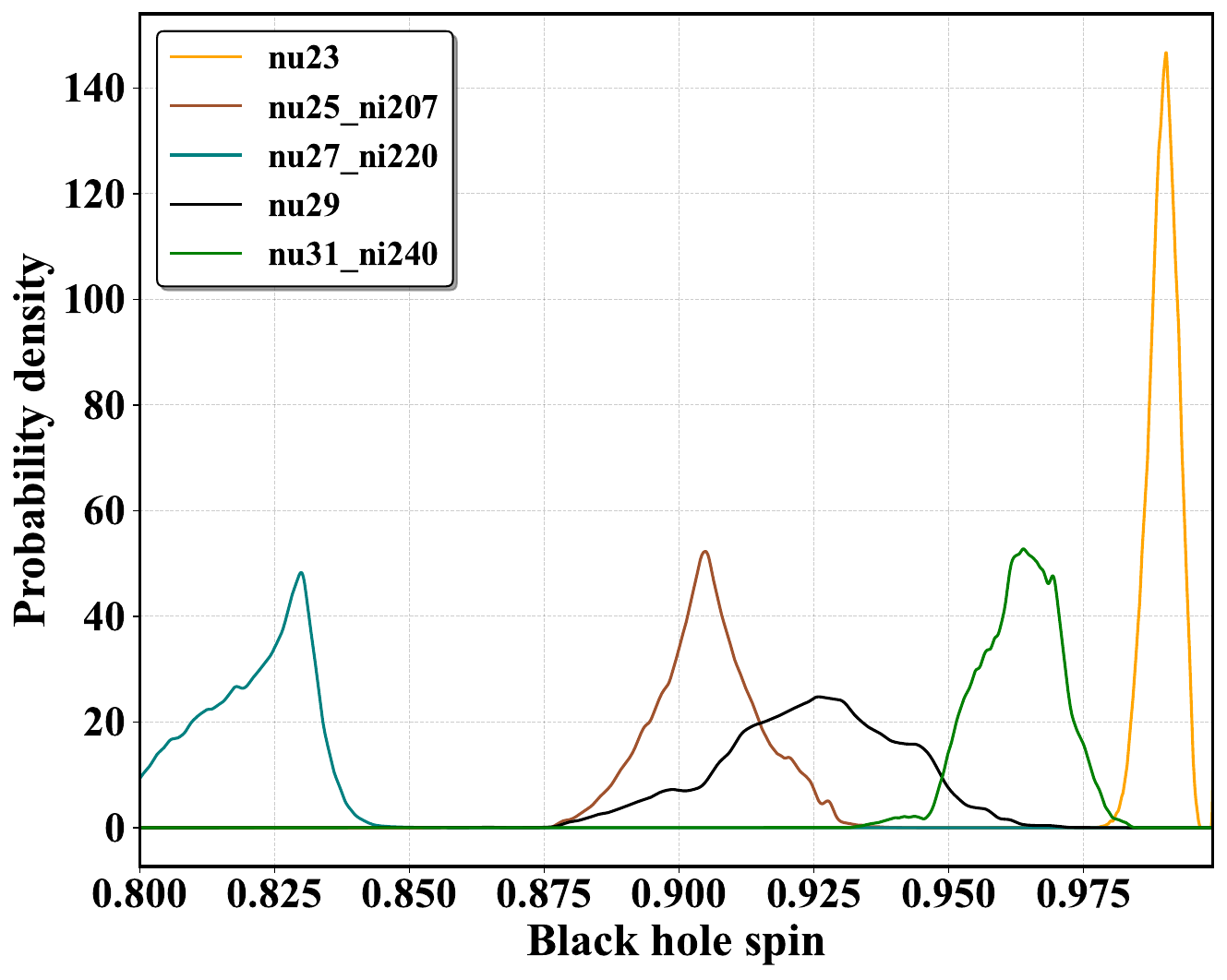}};
\end{tikzpicture}

\caption{\label{figure2} The NuSTAR (top panel) best-fitted spectra along with the respective data-to-model ratios. The bottom panel shows the posterior probability distributions for the black hole spin derived from MCMC analysis. Evidently there is no substantial overlap between successive observational epochs.} 
\end{figure}

All soft-state spectra are fitted using the \texttt{xspec} model: \texttt{tbabs}$*$(\texttt{thcomp}$*$\texttt{kerrbb2}$+$\texttt{bbody}$+$\texttt{bbody}). The second blackbody component is essential for fitting the spectra below 0.5 keV, and is therefore omitted when fitting standalone NuSTAR data. This configuration represents a highly physically motivated model for accreting black holes in the soft state. The \texttt{tbabs} component accounts for galactic absorption, which primarily shapes the continuum within the $0.2\text{--}2.0\text{ keV}$ range. The Comptonized component is handled by the \texttt{thcomp} convolution model \citep{Zdziarski_2020}, applied directly to the seed photon emission from a thin relativistic accretion disk. The disk emission is described by the \texttt{kerrbb2} extension of the \texttt{kerrbb} model that utilizes pre-calculated non-LTE stellar atmosphere tables for the color-correction factor treatment. Employing this framework allows us to robustly constrain both the black hole spin and the fundamental physical properties of the source.

\begin{table*}
\begin{center}
\begin{threeparttable}
\caption{\label{Table3} Best-fit results for MAXI J1820+070 throughout the soft state.}
\begin{tabular}{l l l l l l l}
\hline \\ [0.005ex]  
\multirow{2}{*}{OBSID} & NuSTAR & nu23 & nu25 & nu27 & nu29 & nu31 \\
  & NICER & - & ni207 & ni220 & - & ni240 \\  
\hline \\ [0.01ex]
\multicolumn{7}{c}{Model: \texttt{tbabs}$*$(\texttt{thcomp}$*$\texttt{kerrbb2}$+$\texttt{bbody}$+$\texttt{bbody})} \\
\hline \\ [0.01ex]
\texttt{tbabs} & nH (10$^{22}$ cm$^{-2}$) & 0.1 (fixed) & 0.114$^{+0.003}_{-0.002}$ & 0.117$^{+0.003}_{-0.002}$ & 0.1 (fixed) & 0.109$^{+0.003}_{-0.002}$ \\ [1.5ex]
\texttt{thcomp} & $\Gamma$ & 1.925$^{+0.011}_{-0.021}$ & 1.724$^{+0.022}_{-0.012}$ & 1.717$^{+0.016}_{-0.009}$ & 2.185$^{+0.260}_{-0.020}$ & 2.427$^{+0.042}_{-0.055}$ \\ [1.5ex]
 & kT$_e$ (keV) & 22.99$^{+3.13}_{-1.30}$ & 22.62$^{+6.03}_{-1.92}$ & 23.12$^{+5.10}_{-1.68}$ & 14.67$^{+112.55}_{-1.08}$ & 131.05$^{+14.51}_{-82.49}$ \\ [1.5ex]
 & f$_{sc}$ (\%) & 23.65$^{+0.62}_{-0.68}$ & 5.96$^{+0.30}_{-0.20}$ & 2.16$^{+0.07}_{-0.10}$ & 1.67$^{+1.22}_{-0.10}$ & 3.32$^{+0.12}_{-0.64}$ \\ [1.5ex]
\texttt{kerrbb2} & $a$ & 0.991$^{+0.003}_{-0.006}$ & 0.909$^{+0.013}_{-0.020}$ & 0.832$^{+0.002}_{-0.032}$ & 0.948$^{+0.001}_{-0.054}$ & 0.966$^{+0.008}_{-0.016}$ \\ [1.5ex] 
& M$_{dd}$ (10$^{18}$ g/s) & 0.211$^{+0.032}_{-0.015}$ & 0.359$^{+0.051}_{-0.035}$ & 0.475$^{+0.070}_{-0.008}$ & 0.165$^{+0.094}_{-0.002}$ & 0.118$^{+0.027}_{-0.014}$ \\ [1.5ex]
\texttt{blackbody} & kT (keV) & - & 0.196$^{+0.003}_{-0.004}$ & 0.194$^{+0.003}_{-0.004}$ & - & 0.201$^{+0.003}_{-0.003}$ \\ [1.5ex]
 & norm & - & 0.465$^{+0.014}_{-0.017}$ & 0.405$^{+0.010}_{-0.017}$ & - & 0.381$^{+0.009}_{-0.012}$ \\ [1.5ex]
\texttt{blackbody} & kT (keV) & 0.488$^{+0.003}_{-0.018}$ & 0.478$^{+0.005}_{-0.006}$ & 0.474$^{+0.002}_{-0.005}$ & 0.507$^{+0.0003}_{-0.018}$ & 0.473$^{+0.004}_{-0.005}$ \\ [1.5ex]
 & norm & 0.514$^{+0.132}_{-0.005}$ & 0.584$^{+0.014}_{-0.020}$ & 0.447$^{+0.005}_{-0.022}$ & 0.338 $\pm$ 0.021 & 0.408$^{+0.007}_{-0.015}$ \\ [1.5ex]
 \hline \\ [0.01ex]
 \multicolumn{2}{c}{$l=L/L_{Edd}$ (\%)} & 4.83 & 4.82 & 5.26 & 2.61 & 2.07 \\ [1.5ex]
 \hline \\ [0.01ex]
 \multicolumn{2}{c}{$\chi ^2$/d.o.f} & 581/421 & 446/477 & 561/482 & 222/195 & 387/322 \\ [1.5ex]
 \hline \\ [0.005ex]
 \multicolumn{7}{c}{Low spin best-fit results} \\
 \hline \\ [0.01ex]
 \multicolumn{7}{c}{Model: \texttt{tbabs}$*$(\texttt{thcomp}$*$\texttt{kerrbb2}$+$\texttt{bbody})} \\ 
 \hline \\ [0.01ex]
 \multicolumn{2}{c}{$\chi ^2$/d.o.f} & - & 1222/479 & 1254/484 & - & 1275/324 \\ [1.5ex]
 \hline \\ [0.01ex]
\end{tabular}
\begin{tabular}{l l l l l l l l l}    
\multicolumn{9}{c}{Fixed parameters} \\ 
\hline \\ [0.01ex]
\texttt{thcomp} & \multicolumn{8}{c}{\texttt{kerrbb2}} \\ [1.5ex] 
z & $\eta$ & $i$ ($^\circ$) & M$_{bh}$ (\(M_\odot\)) & D (kpc) & self-irradiation & limb-darken. & $\alpha$ (viscos.) & norm \\ [1.5ex]    
0 & 0 & 63.0 (1) & 8.48 (2) & 2.96 (1) & on & on & 0.01 & 1.0 \\ [1.5ex]    
\hline \\ [0.01ex]
\end{tabular}
\end{threeparttable}
\end{center}
\textbf{References}. (1) \citet{Atri_2020}; (2) \citet{Torres_2020}.
\end{table*} 

Our best-fit results are presented in Table \ref{Table3}, with all parameters being well-constrained by the data. The model is statistically consistent with the observed spectra across all epochs. The slightly elevated residuals observed in specific epochs are primarily driven by unmodeled reflection features that become more prominent following the optimal data grouping. 

The parameter uncertainties presented in Table \ref{Table3} are derived from a Markov Chain Monte Carlo (MCMC) analysis performed on our best-fit models. Specifically, we employed the Goodman-Weare sampler with 20 walkers per free parameter and 12500 steps per walker, discarding the first 20\% of the samples as burn-in. Both trace plots and the Geweke diagnostic confirm chain convergence. Moreover, uniform priors were assumed for all free parameters, with no upper or lower bounds imposed beyond the physical limits of the model. The parameter uncertainties reported correspond to the 90\% credible interval (5th–95th percentile) of the marginalized posterior distribution for each parameter. 

\begin{table}
\begin{center}
\begin{threeparttable}
\caption{\label{Table4} Best-fit results for the nu23 epoch including modelization of the reflected spectrum.}
\begin{tabular}{l l l}
\hline \\ [0.005ex]  
OBSID & NuSTAR & nu23 \\  
\hline \\ [0.01ex]
\multicolumn{3}{c}{Model: \texttt{kerrbb2}$+$\texttt{bbody}$+$\texttt{relxillCp}} \\
\hline \\ [0.01ex]
\texttt{kerrbb2} & $a$ & 0.995$^{+0.002}_{-0.011}$ \\ [1.5ex] 
& M$_{dd}$ (10$^{18}$ g/s) & 0.146$^{+0.043}_{-0.013}$ \\ [1.5ex]
\texttt{blackbody} & kT (keV) & 0.496$^{+0.007}_{-0.010}$ \\ [1.5ex]
 & norm & 0.445$^{+0.007}_{-0.023}$ \\ [1.5ex]
\texttt{relxillCp} & q$_1$ & 1.862$^{+0.171}_{-0.238}$ \\ [1.5ex]
 & $\Gamma$ & 1.941$^{+0.015}_{-0.008}$ \\ [1.5ex]
 & $\log(\xi)$ & 3.10$^{+0.22}_{-0.10}$ \\ [1.5ex] 
 & A$_{Fe}$ & 10.00$^{+0.02}_{-0.97}$ \\ [1.5ex]
 & R$_f$ & 0.230$^{+0.054}_{-0.013}$ \\ [1.5ex]
 & norm & 0.417$^{+0.019}_{-0.004}$ \\ [1.5ex]
 \hline \\ [0.01ex]
 \multicolumn{2}{c}{$\chi ^2$/d.o.f} & 644/456 \\ [1.5ex]
 \hline \\ [0.01ex]
\end{tabular}
\end{threeparttable}
\end{center}
\end{table} 

In order to obtain well-constrained spin estimates, we fix the black hole mass, source distance, and disk inclination to the values reported by \citet{Atri_2020, Torres_2020} for MAXI J1820+070. All free and frozen model parameters, along with the goodness-of-fit metrics, are compiled in Table \ref{Table3}. 

Fig. \ref{figure2} presents our best-fit spectral models for each epoch, along with the corresponding data-to-model residuals. We also show the posterior distributions for the spin parameter independently constrained from each epoch's MCMC analysis. 

\begin{figure}[ht] 
\begin{tikzpicture}
\centering
  \node (img1)  {\includegraphics[width=\linewidth]{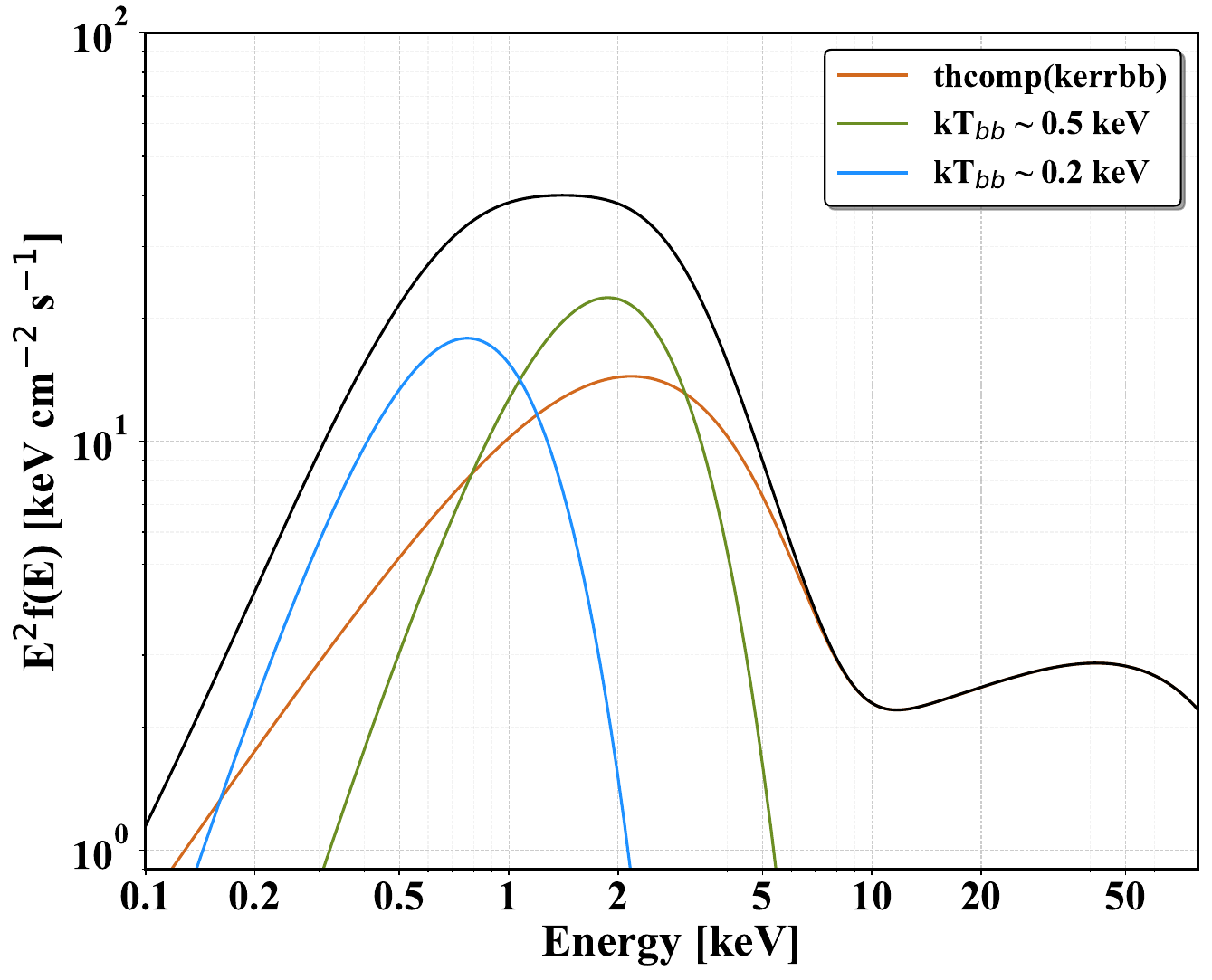}};
\end{tikzpicture}

\caption{\label{contribution} Contribution of each component to the unabsorbed spectrum of epoch nu25. The parameter values employed are the best-fit results of Table \ref{Table3}.}
\end{figure}

The spectra are well-described by the expected multi-color disk emission complemented by two distinct blackbody components at $kT_{\text{bb}} \sim 0.2\text{ keV}$ and $\sim 0.5\text{ keV}$. This spectral decomposition is illustrated in Fig. \ref{contribution}, which displays the contribution of each individual component to the unabsorbed model for the nu25 fit. The soft X-ray flux is entirely reproduced by the combined thermal disk emission and the two blackbody components. Notably, omitting the $0.2\text{ keV}$ blackbody introduces severe systematic residuals below $0.5\text{ keV}$, while omitting the $0.5\text{ keV}$ blackbody causes a substantial data-to-model discrepancy across the $2\text{--}5\text{ keV}$ bandpass.

Moreover, spectral analysis of NICER-only data over the $0.2\text{--}10\text{ keV}$ band, using \texttt{OBSID} 1200120236 (ni236) (obtained after the ni220 epoch), is consistent with the NuSTAR-only fits. More specifically, the best fit for ni236 ($\chi ^2/\text{d.o.f} = 141/153$) yields a black hole spin of $a = 0.938^{+0.032}_{-0.026}$, in broad agreement with epoch nu29. Notably, the \texttt{thcomp} model was removed to avoid overfitting caused by the large number of free model parameters. NICER observation 1200120197 (ni197) has been reported to correspond to the intermediate state and was therefore excluded from our analysis \citep{Wang_2021}.     
	
Evidently, 20\% of the disk emission, that later drops below 5\%, is scattered in the corona, producing a power-law spectrum with a spectral index spanning $\Gamma = 1.9\text{--}2.4$. The temperature of the corona electrons approximates 20 keV across the soft state, which is consistent with previous theoretical and observational studies \citep{Burke_2016, Kajava_2016, Jana_2022}.

For the nu23 epoch, a considerable deviation is observed between the data and the model due to intense reflection features driven by the high count rate. To mitigate this effect and maintain statistical consistency with the remaining epochs, we exclude the $6\text{--}9\text{ keV}$ band from the nu23 results of Table \ref{Table3}, and instead conduct an independent spectral analysis of the iron K complex and prominent reflection features.

To directly test whether this exclusion biases the continuum-derived parameters, we perform a cross-validation on nu23 by modeling the entire NuSTAR bandpass, including the reflected spectrum, with the \texttt{relxillCp} variant of the \texttt{relxill} model family \citep{Dauser_2014, Dauser_2022}. The results are presented in Table \ref{Table4} along with the respective MCMC-based parameter uncertainties. The inclination is fixed at $i=63^\circ$, the corona temperature is fixed at $kT_e=23\text{ keV}$ per Table \ref{Table3}, and the black hole spin is tied between \texttt{kerrbb2} and \texttt{relxillCp}. This choice is warranted because, in this soft-state epoch, reflection features are comparatively weak, so \texttt{relxillCp} alone cannot self-consistently constrain the spin independently of the continuum. Tying the spin to the \texttt{kerrbb2} value is therefore necessary to obtain a well-posed fit, and examine whether the reflection-inclusive model can reproduce the full NuSTAR bandpass without systematically distorting the remaining continuum parameters. 

The resulting emissivity profile collapses to a single, unbroken power law with index $q_1=1.86$, consistent with a compact, centrally illuminating corona. The full-bandpass fit yields $a=0.995^{+0.002}_{-0.011}$, in agreement with $a=0.991^{+0.003}_{-0.006}$ from the continuum-only fit of Table \ref{Table3}. The remaining best-fit parameters of Table \ref{Table4} are likewise consistent with Table \ref{Table3}, with the exception of the mass accretion rate and blackbody normalization, which show a modest decrease, plausibly reflecting a redistribution of flux below 6 keV to the reflected spectrum. This indicates that reflection features within the fitted NuSTAR band do not measurably bias the nu23 continuum-derived spin and blackbody temperature.

\subsection{Source evolution across the soft state}

\begin{figure*}[ht] 
\begin{tikzpicture}[inner sep=0, outer sep=0]
\centering
  \node[node distance=0cm, yshift=1cm] (img1) {\includegraphics[width=0.45\linewidth]{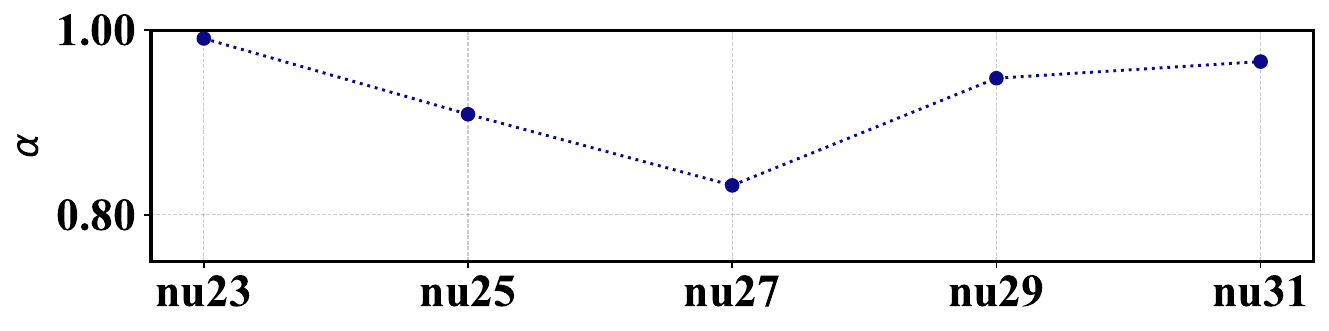}};
  \node[below= of img1, node distance=0cm, yshift=1.35cm] (img2) {\includegraphics[width=0.45\linewidth]{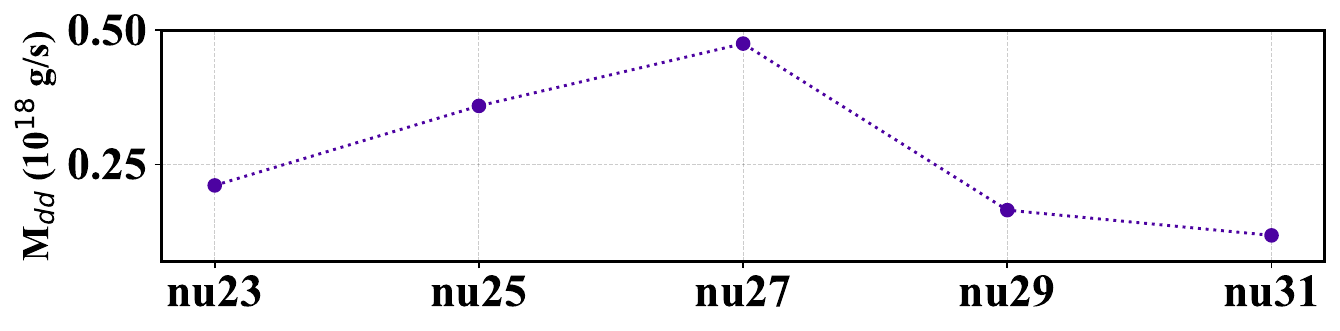}};
  \node[below= of img2, node distance=0cm, yshift=1.34cm] (img3)  {\includegraphics[width=0.45\linewidth]{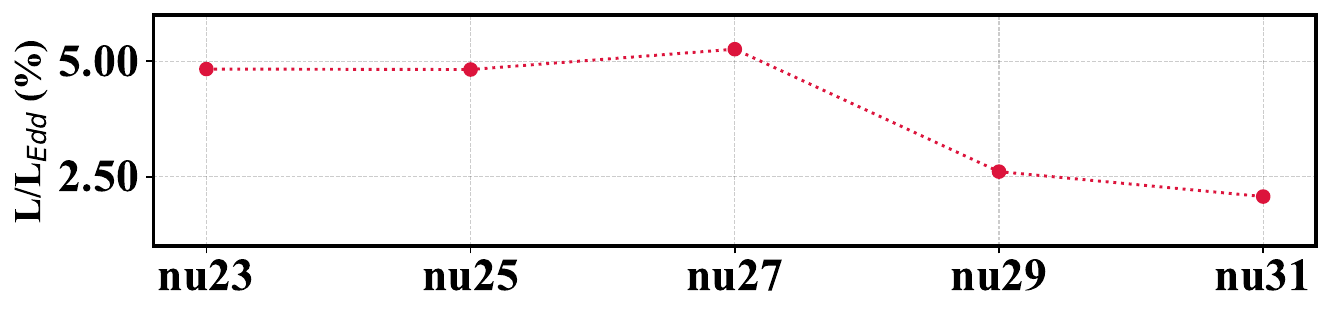}};
  \node[below= of img3, node distance=0cm, yshift=1.35cm] (img4)  {\includegraphics[width=0.45\linewidth]{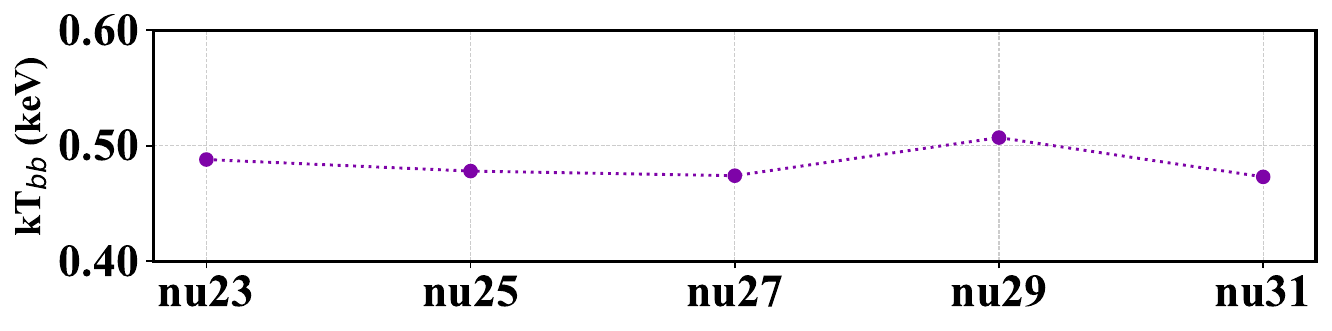}};
  
  \node[right= of img1] (img10)  {\includegraphics[width=0.45\linewidth]{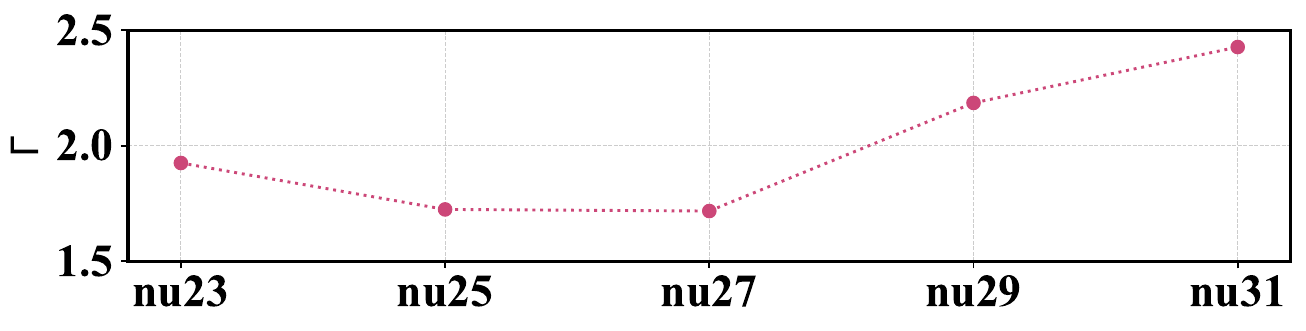}};
  \node[below= of img10, node distance=0cm, yshift=1.347cm] (img11) {\includegraphics[width=0.45\linewidth]{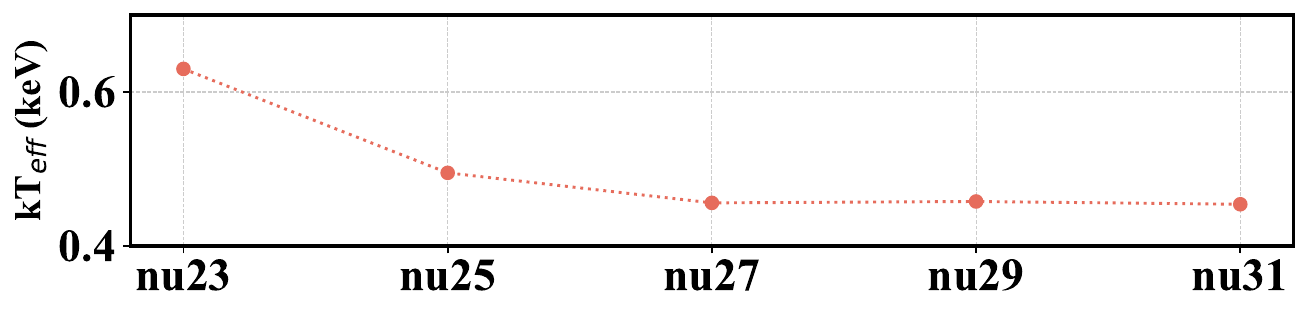}};
  \node[below= of img11, node distance=0cm, yshift=1.35cm] (img12)  {\includegraphics[width=0.45\linewidth]{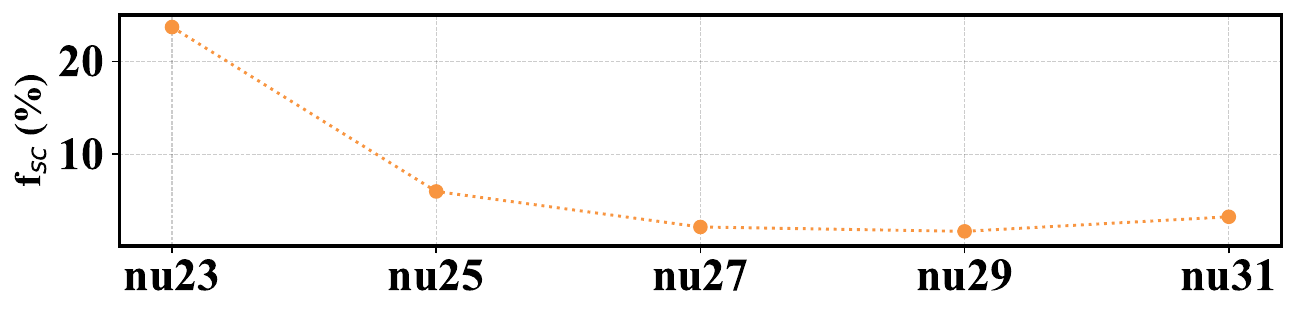}};
  \node[below= of img12, node distance=0cm, yshift=1.355cm] (img13)  {\includegraphics[width=0.45\linewidth]{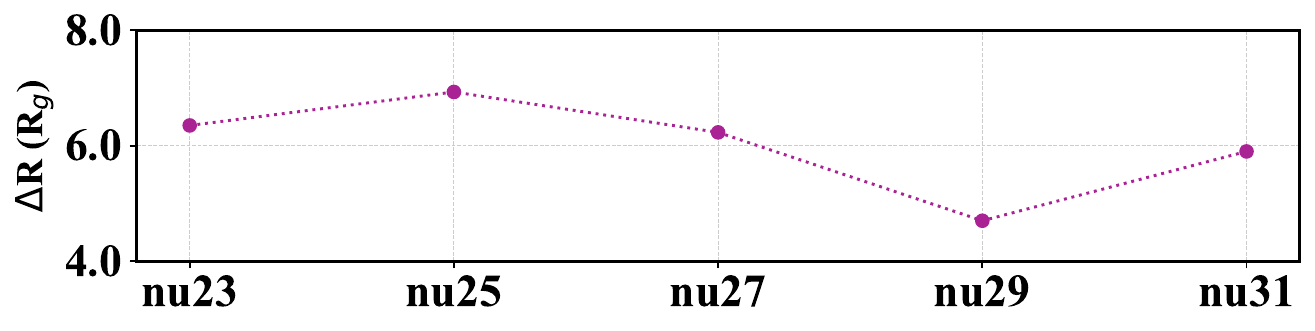}};
\end{tikzpicture}

\caption{\label{figure4} Source evolution during the soft state based on the variation of the best-fit model parameters across all epochs. The last plot refers to the radius of the blackbody emission region with kT$_{bb}\approx 0.5\text{ keV}$. In addition, $kT_{\text{eff}}$ refers to the maximum effective temperature of the accretion disk.}
\end{figure*}

Fig. \ref{figure4} illustrates the source evolution throughout the soft state, as inferred from the variance of our best-fit model parameters. It is evident that despite variations in the estimated spin and accretion rate parameters across different epochs, the Eddington luminosity ratio remains remarkably stable until epoch nu27. This specific epoch marks the onset of a gradual transition in the source's accretion state, following a decrease in luminosity of approximately 50\%.

The variation in the disk luminosity has a profound effect on the corona dynamics. The substantial increase in high-energy seed photons from the inner disk, following its extension closer to the black hole, enhances Compton cooling. This results in spectral softening beyond epoch nu27, tracking a spectral index increase to $\Gamma = 2.2\text{–}2.5$.

The blackbody emission component remains remarkably steady at $kT \approx 0.5\text{ keV}$ across all epochs, displaying a maximum variation of only 7\% between nu27 and nu29. This minor fluctuation is likely an artifact of tighter parameter constraints afforded by the inclusion of the NICER data. The stability of this temperature suggests that the blackbody-emitting region is physically distinct from both the main accretion disk and the variable corona. Nonetheless, we will be able to delve deeper into the origin of this emission once we investigate how the temperature profile of the disk evolves throughout the soft state. 

Further investigation prompts us to calculate the physical size of the emitting region using the blackbody normalization constant and track its evolution throughout the soft state. As shown in the bottom right panel of Fig. \ref{figure4}, this region varies between $4.5\text{--}6.5R_{\text{g}}$ ($\sim 56\text{--}82\text{ km}$), values that are far too large to be associated with the plunging region, especially given that the black hole spin is constrained to $a > 0.8$. Moreover, the structural evolution of this emission region mirrors the corona variation patterns, hinting at a potential physical connection between the two components. Nevertheless, a dynamic coupling of this nature should be further investigated with the inclusion of a larger ensemble of NICER/NuSTAR data to yield even tighter parameter constraints.

\section{Spin implications and comparison with previous estimates}\label{Sec4}

Our spectral analysis indicates a high dimensionless black hole spin of $a\approx$ 0.9. This result contrasts with previous continuum-fitting measurements below 25 keV using NICER and Insight-HXMT observations \citep{Zhao_2021, Sai_2024} and timing analysis of QPO frequencies \citep{Fiori_2025}. However, it is broadly consistent with measurements obtained through alternative spin-determination techniques \citep{Bhargava_2021, Draghis_2023}.

Continuum-fitting analyses of the soft state in MAXI J1820+070 using Insight-HXMT and NICER data indicate considerably lower spins between $a$ = 0.1–0.3 \citep{Zhao_2021, Guan_2021, Sai_2024}. In the first case, the spectra were extracted from the $2\text{--}10\text{ keV}$ band of the Low Energy (LE) HXMT instrument and the $10\text{--}25\text{ keV}$ band of the Medium Energy (ME) telescope. NICER data cover the $1\text{--}10\text{ keV}$ energy band.  

In the majority of studies supporting lower black hole spin values, the soft X-ray spectra were modeled using variations of the standard \texttt{tbabs}$*$\texttt{simpl}$*$\texttt{kerrbb} model. However, these spectral fits frequently omit low-energy data below 0.5 keV or even 0.8 keV, exhibit noticeable residual deviations and large spin parameter uncertainties, or require systematic errors of 1–2\% to achieve statistically acceptable fits. Under the assumption of low spin, \citet{Fabian_2020} showed that an additional blackbody component with $kT\approx 1\text{ keV}$ is necessary to describe the NuSTAR spectra below 10 keV. They report that this excess emission is also confirmed by NICER data between $0.8\text{--}7\text{ keV}$ after imposing a 1.5\% systematic error. However, an inclination near $70^{\circ}$ yields the most negative (retrograde) spin solutions.

In the present work, we jointly fit simultaneous NICER and NuSTAR observations of MAXI J1820+070 across the full $0.2\text{--}79\text{ keV}$ broadband spectrum. Our modeling framework self-consistently accounts for relativistic color-correction effects and thermal Comptonization. Nevertheless, the primary departure from previous low-spin estimations is our inclusion of the two soft blackbody components. Accounting for this low-energy excess yields statistically significant agreement with the joint spectra (see Table \ref{Table3}) and drives the inferred black hole spin toward higher values ($a > 0.8$).

\subsection{High-spin vs. Low-spin black hole}

To resolve the conflict between the high- and low-spin scenarios, we compare the two cases by refitting the joint NICER/NuSTAR spectra using the \texttt{tbabs}$*$(\texttt{thcomp}$*$\texttt{kerrbb2}$+$\texttt{bbody}) model, adopting initial values of $a = 0.2$ and $kT = 1\text{ keV}$. Similar to our high-spin spectral fits, we utilize the $0.2\text{--}5\text{ keV}$ NICER and $5\text{--}79\text{ keV}$ NuSTAR bandpasses, with all model parameters tied across the instruments. We note that the remaining best-fit parameters are consistent between both cases.  

\begin{figure*}[ht] 
\begin{tikzpicture}
\centering
  \node (img1)  {\includegraphics[width=0.45\linewidth]{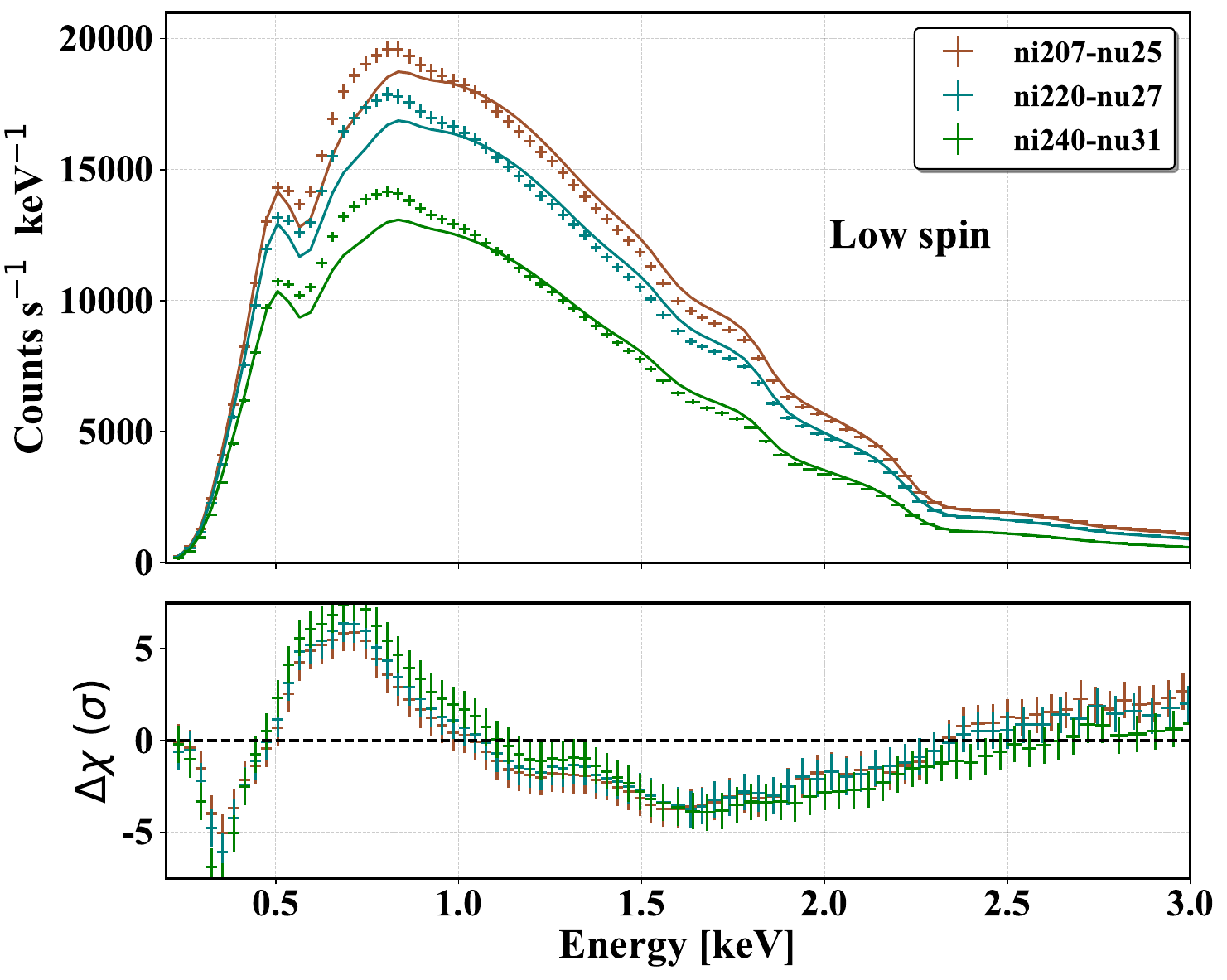}};
  \node[right= of img1] (img2)  {\includegraphics[width=0.45\linewidth]{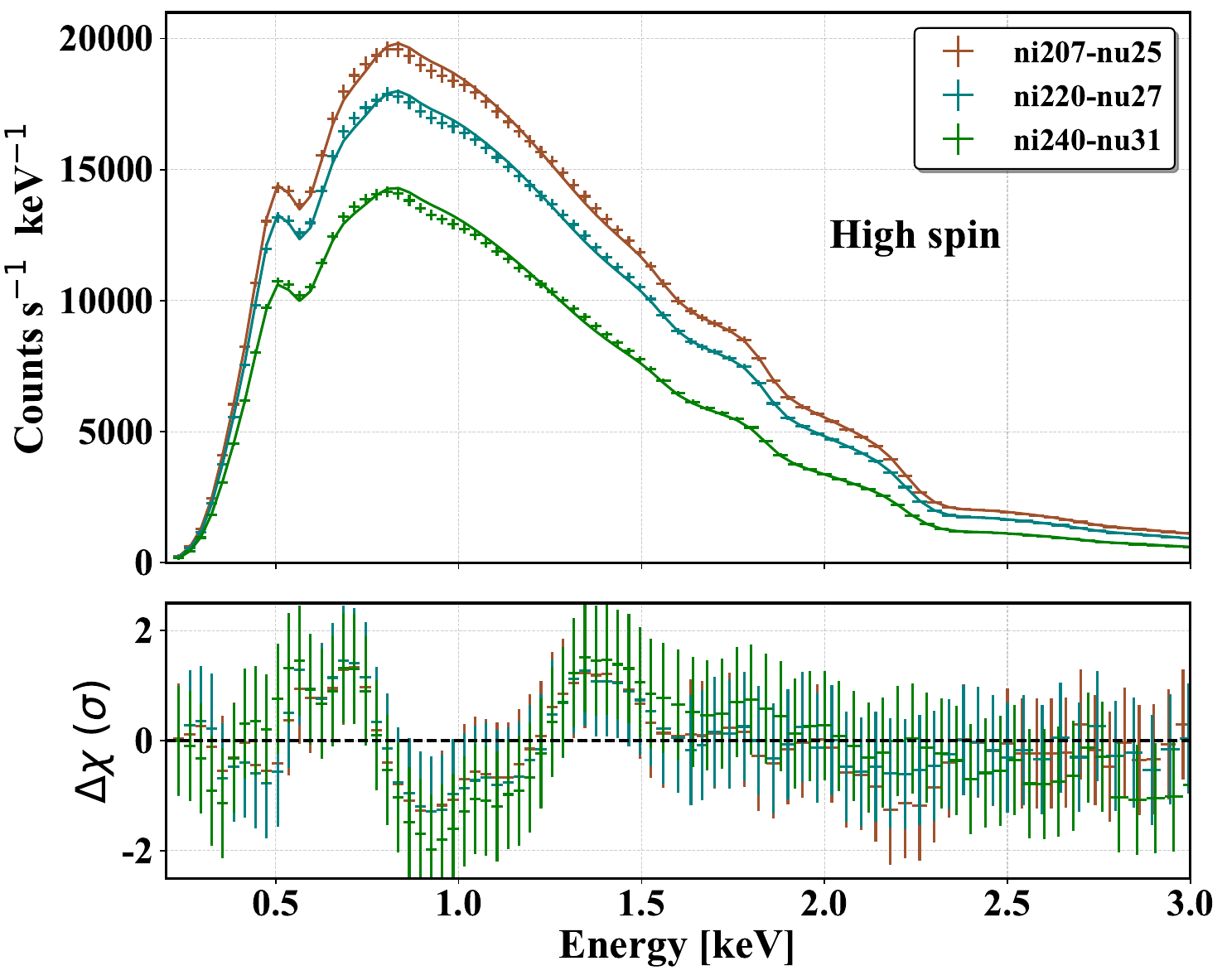}};
\end{tikzpicture}

\caption{\label{figure3} The low-spin (left panel) and the high-spin (right panel) best-fits of the joint NICER/NuSTAR spectra along with their respective residuals. The low-spin fit is associated with a blackbody component with $kT\approx 1\text{ keV}$, while the high-spin fit with two blackbody components with $kT\approx 0.2\text{ keV}$ and $kT\approx 0.5\text{ keV}$, respectively.}
\end{figure*}

Fig. \ref{figure3} shows the direct comparison between the two distinct cases below 3 keV, where the bulk of the disk continuum resides. It is evident that, while the low-spin model produces seemingly reasonable fits above 1 keV, there are significant residuals in the softer part of the spectrum, alluding to an unmodeled excess emission. This emission is consistently described by the $kT\approx 0.2\text{ keV}$ blackbody in the high-spin model. Notably, this excess extends even to the $0.5\text{--}1\text{ keV}$ range, which is considered highly reliable within the NICER bandpass. The goodness-of-fit statistics for the high-spin ($\chi ^2/\text{d.o.f} = 446/477$, $\chi ^2/\text{d.o.f} = 561/482$, and $\chi ^2/\text{d.o.f} =387/322$) and low-spin ($\chi ^2/\text{d.o.f} = 1222/479$, $\chi ^2/\text{d.o.f} = 1254/484$, and $\chi ^2/\text{d.o.f} = 1275/324$) models are listed in Table \ref{Table3}.

Another problematic aspect of the low-spin hypothesis is that the best-fit spin converges to a negative value ($a = -0.38$). This result is expected from previous studies given the source's inclination of $63^{\circ}$. Physically, this behavior arises because the model attempts to compensate for the excess emission by artificially increasing the mass accretion rate and reducing the spin. Freezing the spin parameter at moderately low values ($a\approx 0.2$), substantially worsens the fit quality, with $\chi ^2/\text{d.o.f}$ increasing from $1275/324$ to $3444/325$ for the ni240-nu31 epoch.

On the other hand, the high-spin model is statistically consistent with the data across the entire disk emission profile, including the soft energies below 0.5 keV. The systematic residual ($\Delta\chi$) structures observed near 0.5 keV, 1.5 keV, and 2.2 keV are well-known instrumental artifacts, corresponding to the carbon, oxygen, and aluminum K-edges, as well as the gold M-absorption edge features originating from the detector windows and filters \citep{Prigozhin_2016}.

The NICER spectrum below $0.5\text{ keV}$ is often deemed unreliable and is frequently omitted in spectral analyses due to instrumental calibration uncertainties. Nonetheless, our results demonstrate that these low-energy data can be robustly and self-consistently modeled. To evaluate whether the high-spin model is simply absorbing calibration residuals, we refit the joint spectra while gradually shrinking the energy range to $1\text{--}79\text{ keV}$, confirming that the best-fit model parameters remain consistent. 

In addition, 2D posterior probability contours derived from our MCMC analysis (see Fig. \ref{prob_contour} of Appendix \ref{Append}) verify that there are no significant degeneracies between the parameters affecting the spectra below 0.5 keV, specifically the nH column density of the \texttt{tbabs} model and the cooler blackbody normalization. Fig. \ref{figure3} clearly shows that the additional blackbody is necessary for fitting the data within the broader $0.2\text{--}1.0\text{ keV}$ band, rather than just the range below 0.5 keV. Consequently, fitting the joint ni207-nu25 spectra with the low-spin model in the restricted $1\text{--}79\text{ keV}$ range yields an improved but still unacceptable fit ($\chi ^2/\text{d.o.f} = 731/450$) and an increased, but still negative, spin of $a = -0.17$.  

Standalone NuSTAR data above 3 keV, grouped with \texttt{grppha} to a minimum of 25 counts per bin, are described equally well by both the high- and low-spin models. However, using the optimal binning method of \citet{Kaastra_2016} facilitates a clear distinction between them. Under this optimal binning scheme, the high-spin model is statistically preferred. Specifically, for epoch nu29, the high-spin model yields a fit of $\chi ^2/\text{d.o.f} = 222/194$ compared to $251/195$ for the low-spin scenario. Similarly, for epoch nu31 (excluding the NICER data), the high-spin model yields $\chi ^2/\text{d.o.f} = 336/238$ compared to $402/239$ for the low-spin case. In these optimally binned fits, the low-spin models yield values of $a = 0.18\text{--}0.2$. In the case of nu23, there is no reliable statistical distinction between spin scenarios: the low-spin solution ($a = -0.24$, $\chi ^2/\text{d.o.f} = 586/421$) is statistically comparable to the high-spin solution ($\chi ^2/\text{d.o.f} = 581/421$). This could be attributed to the exclusion of the $6\text{--}9\text{ keV}$ band. However, fixing the spin to $a = 0.2$, as favored by \citet{Fabian_2020} and \citet{Zhao_2021}, significantly degrades the fit ($\chi ^2/\text{d.o.f} = 758/422$).           

Moreover, our results, indicating a high spin of $a > 0.8$, agrees with the QPO frequency evolution derived from NICER power density spectra during the hard state. Application of the relativistic precession model (RPM) gives a spin of $a = 0.799^{+0.016}_{-0.015}$ \citep{Bhargava_2021}. Relativistic reflection spectroscopy of multi-epoch NuSTAR data across both primary spectral states indicates a highly rotating black hole with $a = 0.988^{+0.006}_{-0.028}$ and a disk inclination of $\theta = 64^{+3}_{-4}$ \citep{Draghis_2023}. This inclination is also consistent with the jet angle determined from radio parallax measurements \citep{Atri_2020}, providing further support for the high-spin estimate.

\section{Indications of a small disk truncation}\label{Sec5}

The black hole spin seems to exhibit variations throughout the soft state, decreasing by up to 16\% and reaching a minimum of $a\approx 0.83$. This evolution, as shown in Fig. \ref{figure4}, does not appear to be driven by statistical bias. This is supported by the posterior distributions in Fig. \ref{figure2}, which show no substantial overlap between successive observational epochs, indicating that the apparent spin variation cannot be explained by parameter correlations or random fluctuations in the parameter space. 

The inferred spin is initially very high, consistent with a near-maximally rotating black hole. Approximately halfway through the soft state, the spin decreases, before increasing again a few days later during epoch nu29. However, since the black hole spin is not expected to vary, especially over the month-long timescale covered by the NICER/NuSTAR observations, we instead investigate whether the apparent spin variability reflects changes in other physical properties of the system. In the \texttt{kerrbb2} model, the spin parameter is effectively determined by the ISCO radius. Therefore, variations in the estimated spin correspond to changes in the inner disk boundary and its proximity to the event horizon.

\begin{figure}[ht] 
\begin{tikzpicture}
\centering
  \node (img1)  {\includegraphics[width=\linewidth]{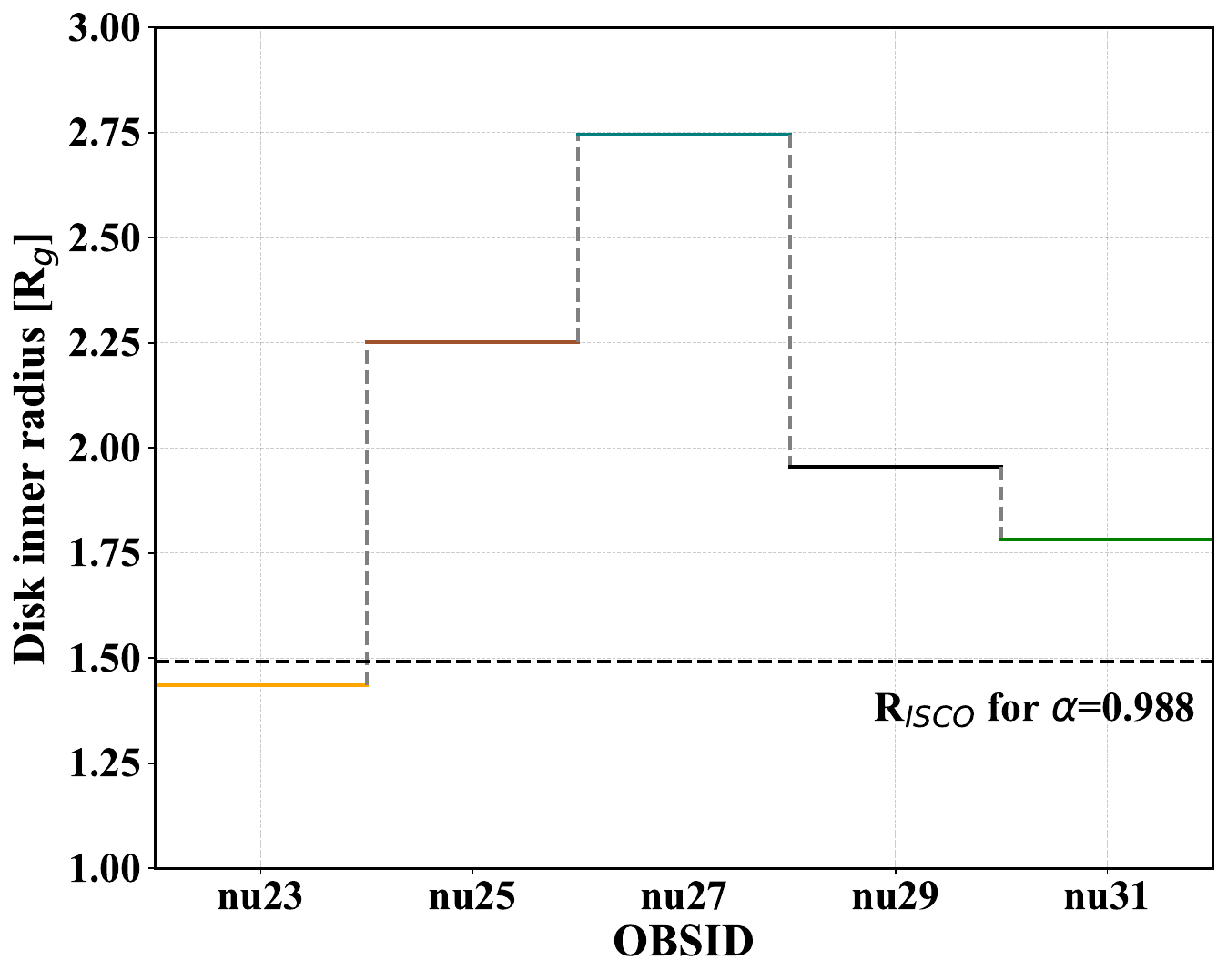}};
\end{tikzpicture}

\caption{\label{figure7} The evolution of the inner disk radius according to the spin estimations of our best-fit model over epochs nu23-nu31. For comparison, the ISCO radius corresponding to $a = 0.988$ from \citet{Draghis_2023} is also plotted.}
\end{figure}

Fig. \ref{figure7} shows the evolution of the inner disk radius across all epochs, as inferred from the best-fit spin parameter of Table \ref{Table3}. The near-maximal prograde spin measured in epoch nu23 corresponds to an accretion disk extending close to the gravitational radius. Subsequently, the disk radius increases to $2.25R_{\text{g}}$ and then to $2.75R_{\text{g}}$ before moving inward again toward the black hole. For comparison, we also plot the ISCO radius corresponding to $a = 0.988$ \citep{Draghis_2023}. Adopting this spin, the offset between the horizontal reference line and the inferred inner disk radii in Fig. \ref{figure7} suggests that the accretion disk might not consistently extend to the ISCO, but instead undergoes slight truncation during the soft state.   

Nevertheless, the aforementioned potential structural changes in the disk could instead reflect intrinsic modifications in the physical state of the accreting gas. Previous studies have supported that spectral evolution between state transitions could be attributed to color correction instead of the commonly speculated disk truncation \citep{Salvesen_2013, Zdziarski_2022}. In the innermost regions, the accretion flow can reach temperatures between $10^5\text{--}10^7\text{ K}$, leading to strong ionization and a regime in which electron scattering dominates over thermalization interactions. This results in spectral hardening, which is accounted for, in the multi-color disk framework, through the color correction factor, $f_{\text{col}} = T_{\text{col}}/T_{\text{eff}} = 1.2-1.9$ (denoted as \texttt{hd} in the \texttt{kerrbb} model) \citep{Li_2005}. 

Our analysis incorporates the \texttt{kerrbb2} model, which accounts for non-blackbody atmospheric effects via a pre-computed table of spectral hardening values derived from fitting \texttt{bhspec} model spectra with \texttt{kerrbb} at fixed input coordinates. We find that the color-correction factor varies within a narrow range of $f_{\text{col}} \approx 1.54\text{--}1.58$ across the soft state. The stability of this parameter implies that variations in non-blackbody atmospheric effects are too minor to serve as a viable explanation for the observed spectral variability.     

\subsection{Evolution of the disk temperature profile across the soft state}

It is evident that an abrupt drop in the temperature of the inner disk regions occurs midway through the soft state (see $kT_{\text{eff}}$ plot in Fig. \ref{figure4}), which cannot be explained solely by variations in the accretion rate. Further examination motivates us to investigate how the disk temperature profile evolves across all epochs. However, the \texttt{kerrbb2} model does not provide direct access to this quantity, since $T_{\text{in}}$ is not an explicit model parameter, unlike the \texttt{diskbb} model.

In \citet{Papavasileiou_2025}, we adopted the pseudo-Newtonian gravitational potential introduced by \citet{Artemova_1996} to derive a spin-dependent temperature–radius relation for accretion disks around rotating black holes. The resulting temperature profile is written as \citep{Papavasileiou_2025}

\begin{equation}
\label{temp_A}
T_A(r)= T_0\left(\frac{(3-\beta )(r-r_h)+\beta r}{3r^{3-\beta}(r-r_h)^{\beta +1}}\right)^{1/4}\left(1-\mathcal{A}\right)^{1/4} ,   
\end{equation}
where $r$ is the radius normalized by the gravitational radius, $r_h= 1+\sqrt{1-a^2}$ is the event horizon depending on the spin, and $\beta = r_{in}/r_h-1$ \citep{Artemova_1996}. The quantity within the brackets is given by
\begin{equation}
\mathcal{A} = \left(\frac{r_{in}}{r}\right)^{(\beta +1)/2}\left(\frac{r-r_{h}}{r_{in}-r_{h}}\right)^{\beta /2} .
\end{equation}
The disk temperature scales as
\begin{equation}
T_0= \left(\frac{3GM_{bh}\dot{M}_{PN}}{8\pi\sigma R_{g}^3}\right)^{1/4} .
\end{equation}     

\begin{figure}[ht] 
\begin{tikzpicture}
\centering
  \node (img1)  {\includegraphics[width=\linewidth]{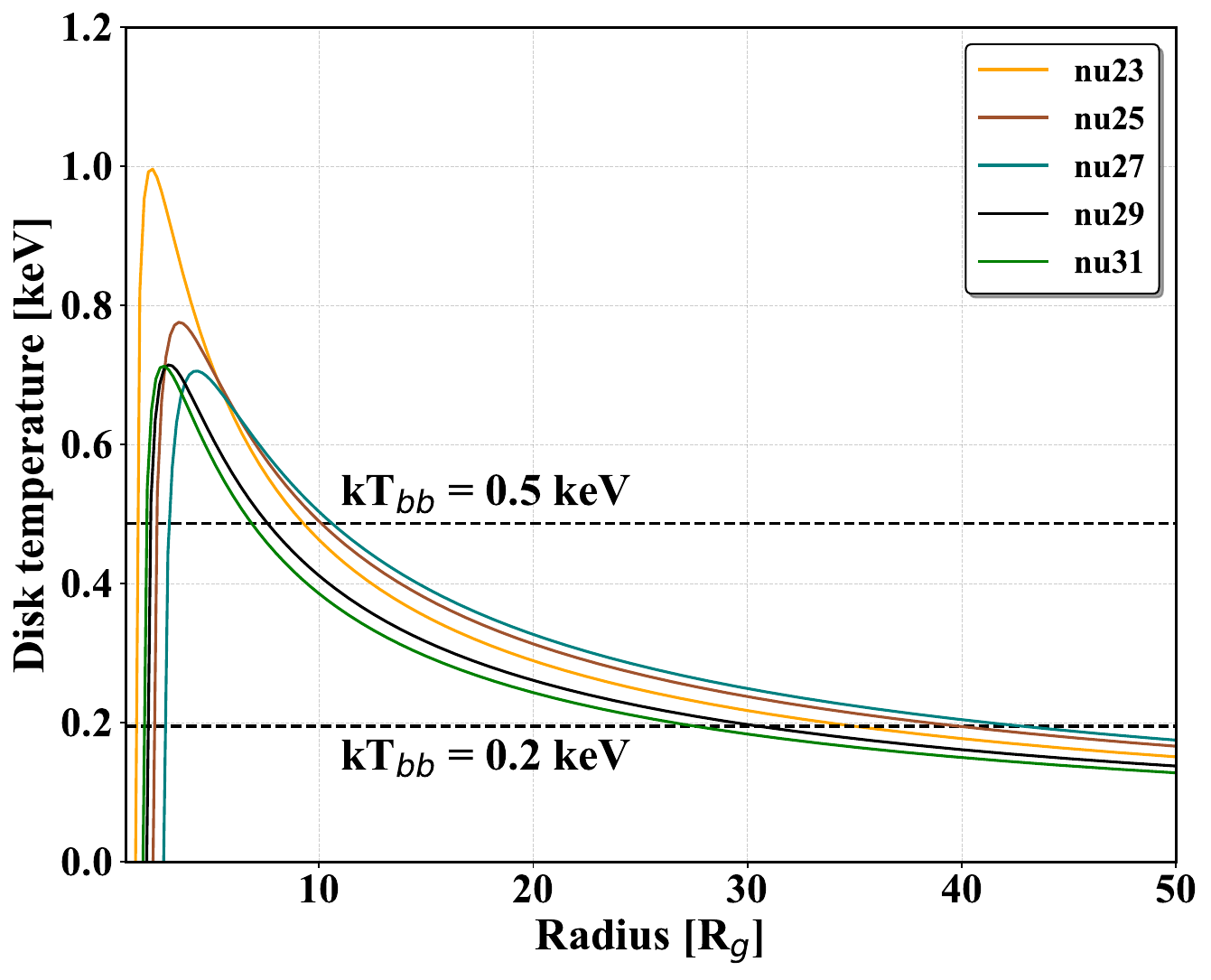}};
\end{tikzpicture}

\caption{\label{figure9} Disk temperature evolution across the soft state. The two blackbody temperatures are also plotted, indicating cooler emission sources than the inner disk regions.}
\end{figure}

Fig. \ref{figure9} displays the color-corrected radial disk temperature profiles derived from Eq. \ref{temp_A} for each observational epoch. Notably, the mass accretion rate was minorly adjusted as $\dot{M}_{PN} = (l/l_{PN})\dot{M}$ with $l_{PN}$ the disk luminosity derived by employing Eq. \ref{temp_A} to absorb any subtle discrepancies between the pseudo-Newtonian treatment and the empirical luminosity profiles from our best-fit analysis. Therefore, the accretion rate correction ranges between $6\text{--}9$\% across all epochs. Initially, the disk is notably hotter, with inner temperatures reaching $\sim 1\text{ keV}$ ($\approx 10^7\text{ K}$), while a larger fraction of its emission is scattered by the corona. Subsequently, the inner disk regions cool abruptly to approximately $0.7\text{ keV}$ leading up to epoch nu27, at which point the disk temperature remains quite stable before the soft state gradually decays.

Table \ref{Table3} and Fig. \ref{figure4} reveal that the luminosity remains fairly stable up to epoch nu27 at $L \approx 0.05\,L_{\text{Edd}}$, whereas the maximum disk temperature varies considerably during this same interval. Given that the color-correction factor varies only marginally and can be approximated as $f_{\text{col}} \approx 1.56$, this behavior represents a clear deviation from the expected $L \propto T^4$ relation. Consequently, this mismatch implies that the inner disk radius does not remain stationary throughout the soft state.  

Similar spectral variability, indicating low-temperature disk emission despite high luminosities, is observed in the very high states of black hole binaries. In these cases, it was speculated that an optically thick, extended corona partially obscured the inner disk. However, explaining the observed spectra required a substantial increase in the inner disk radius. An alternative explanation involves a disk–corona coupling model, which reproduces the observed features more accurately \citep{Done_2006}. Even so, a slight increase in the inner disk radius remains necessary to account for the observations, supporting our claim of minor disk truncation. In this scenario, the corona must be optically thick and extend over a large portion of the innermost, hotter regions of the disk, potentially linking it to the blackbody excess observed below 10 keV.

In a recent study, \citet{Hankla_2025} investigated the structure of the inner regions of weakly magnetized, geometrically thin accretion disks around rapidly spinning black holes. Utilizing three-dimensional GRMHD simulations that incorporate both electron and proton thermodynamic schemes, they demonstrated that the plunging region exhibits a two-temperature structure. Furthermore, they showed that this region extends to radii beyond the ISCO, even at the high mass accretion rates characteristic of the soft state. Specifically, Fig. 11 of \citet{Hankla_2025}  indicates that for a black hole spin of $a = 0.9375$ and an accretion rate of $\dot{m} \equiv \dot{M}/\dot{M}_{\text{Edd}} \approx 10^{-2}$, the thin disk truncates at a radius of approximately $3.5\text{--}4.0R_{\text{g}}$, with this truncation radius shifting inward at higher accretion rates. This numerical result is consistent with our spectral analysis, which indicates a disk truncation radius of $2.75R_{\text{g}}$ at $\dot{m} \approx 0.05$ for a comparable black hole spin.     

\section{The origin of the dual blackbody soft excess}\label{Sec6} 

\citet{Fabian_2020}, analyzing the 2018 NuSTAR dataset, concluded that the soft excess originated near the edge of the plunging region. Their spectral fitting with the \texttt{diskbb}$+$\texttt{bbody}$+$\texttt{cutoffpl} model revealed a blackbody excess at $k T \approx 1\text{ keV}$, whereas the inner disk temperature was found to be only $0.6\text{ keV}$. During this fitting, an inner disk boundary at $5R_{g}$ was assumed, corresponding to a spin of $a\approx 0.2$, and an inclination of $i = 34^{\circ}$, based on the hard-state spectral analysis of \citet{Buisson_2019}. However, these assumptions conflict with the more reliable inclination of $63^{\circ}$ reported by \citet{Atri_2020}, as well as arguments supporting a rapidly spinning black hole. Furthermore, the GRMHD simulations of \citet{Hankla_2025} demonstrate that emission originating within the plunging region is optically thin and predominantly non-thermal due to electron-proton decoupling.

As a continuation of \citet{Fabian_2020}, \citet{Mummery_2024} modeled the blackbody-shaped excess emission between $6\text{--}10\text{ keV}$ using their newly developed \texttt{fullkerr} model, which accounts for emission within the ISCO. Their work demonstrates that spectral fits, accounting for this emission, degrade significantly at higher spin values owing to the contraction of the plunging region, which suppresses the so-called intra-ISCO emission. This behavior aligns with our spectral analysis if we neglect the $k T_{\text{bb}} \approx 0.5\text{ keV}$ blackbody component. A high-spin configuration paired with this additional $0.5\text{ keV}$ blackbody-like spectrum (characterized by a large normalization constant) yields NuSTAR fits that are statistically indistinguishable, at least for the case of non-optimally binned data, from a low-spin framework paired with a hotter, spatially smaller blackbody-like intra-ISCO component. Indeed, \citet{Mummery_2025} explicitly show the mathematical degeneracy between X-ray spectra emerging from a rapidly rotating black hole lacking intra-ISCO emission and a low-spin black hole combined with substantial emission from the plunging region. 

Consequently, jointly fitting the NuSTAR data with simultaneous NICER spectra across the $0.2\text{--}3\text{ keV}$ band is essential to distinguish between these two scenarios, as seen in Fig. \ref{figure3}. To this end, our spectral analysis demonstrates that a high-spin black hole paired with two cooler, spatially larger blackbody components provides a self-consistent description of the X-ray spectra from MAXI J1820+070 across the full NICER/NuSTAR bandpass. Notably, the data were optimally binned and no systematic errors were introduced during the fitting process. This high-spin conclusion is strongly supported by independent relativistic reflection spectroscopy \citep{Draghis_2023}, while our findings regarding a slight disk truncation are entirely consistent with modern GRMHD simulations \citep{Hankla_2025}. Consequently, our constraints on the blackbody-like excess components are highly robust.

Regarding the second blackbody excess found by incorporating the NICER data below $0.5\text{ keV}$, the corresponding emitting region is substantially more extended, spanning $\Delta R \sim 35R_{\text{g}}$ ($\approx 440\text{ km}$). Its low temperature of $k T_{\text{bb}} \approx 0.2\text{ keV}$ closely tracks the outer thermal plateau reached by the disk beyond $\sim 40R_{\text{g}}$, as illustrated in Fig. \ref{figure9}. The smaller warm temperature and the larger size compared to the $0.5\text{ keV}$ blackbody indicate that the emission source is potentially located even further away from the black hole. Furthermore, the structural similarities between these two distinct spectral components point toward a shared physical mechanism. Nonetheless, systematic analysis of a wider pool of NICER/NuSTAR datasets is required to more deeply probe the long-term temporal evolution of this cooler excess.

A soft X-ray excess near 1 keV has been consistently reported in various active galactic nuclei (AGNs). This emission is commonly modeled as a blackbody, although its purely thermal origin is disputed due to its constant temperature that does not scale with the black hole mass and luminosity. Consequently, the soft excess is often attributed either to ionized reflection from the inner disk surface or to thermal Comptonization by a warm corona above the disk \citep{Petrucci_2018, Kubota_2018, Ding_2021}. The latter is the most widely accepted explanation, assuming a warm corona with temperatures between $kT\approx 0.1-1\text{ keV}$ and optical depths of $\tau \approx 10\text{--}40$. In this scenario, the corona reprocesses the UV/optical emission from the inner disk, producing a quasi-thermal excess in the soft X-ray band \citep{Petrucci_2018, Kubota_2018, Ding_2021}.

Both blackbody excess emissions in MAXI J1820+070 span the $0.1\text{--}5\text{ keV}$ energy range and maintain nearly constant temperatures of 0.2 keV and 0.5 keV (see Fig. \ref{figure9}), independent of the source luminosity. Furthermore, their respective fractional luminosities are comparable to the main disk emission, closely mirroring the energetic profiles observed in AGN soft excesses. This behavior strongly implies that a scaling mechanism analogous to that in AGNs drives the soft excess in X-ray binaries. A key difference is that the inner disk in XRBs is considerably hotter, potentially leading to Compton cooling rather than thermal Comptonization. Consequently, the seed photons must originate from much cooler, outer disk regions characterized by ambient temperatures below $0.5\text{ keV}$ and $0.2\text{ keV}$, respectively. As illustrated in Fig. \ref{figure9}, these temperatures correspond to radii beyond $\sim 10R_{\text{g}}$ and $\sim 40R_{\text{g}}$ from the black hole. These large emitting regions are fully consistent with warm Comptonization paradigms, which dictate that warm coronal layers must cover an extended portion of the disk surface \citep{Petrucci_2018}. 

To test the assumption of a warm Comptonization origin, we fit the NICER epoch ni236 by replacing the $k T_{\text{bb}} \approx 0.5\text{ keV}$ blackbody with a power-law with an exponential cutoff. To reduce the number of free parameters and maintain consistency, the nH column density and the $k T_{\text{bb}} \approx 0.2\text{ keV}$ blackbody parameters are fixed to values consistent with the best-fit results of Table \ref{Table3}. This model provides a statistically consistent fit to the data ($\chi ^2/\text{d.o.f} = 154/158$), yielding a black hole spin value near $a = 0.95$. However, the resulting spectral photon index is negative and the high-energy cutoff lies between $0.5\text{--}1.0\text{ keV}$. This physical behavior is consistent with a saturated cooler corona of high optical depth, which produces a Wien-like spectrum rather than the typical power-law Comptonized tail. We stress that this cutoff power-law fit serves only as a phenomenological consistency check on the warm corona scenario and should not be interpreted as an actual Comptonization model. Proper validation would require dedicated Comptonization models accommodating relatively cool electron temperatures.         


\section{Conclusions}\label{Sec7}

The spin of the black hole in MAXI J1820+070 remains a persistent and debated issue. Continuum-fitting of the source’s soft-state spectra generally indicates a low spin ($a < 0.3$), whereas alternative methods, including reflection spectral analysis and the relativistic precession model, favor a higher spin ($a > 0.75$).

We aim to resolve this debate by jointly fitting NICER/NuSTAR spectra over the entire $0.2\text{--}79\text{ keV}$ bandpass, obtained while the source was in the soft state during the summer of 2018. Our results strongly suggest that the soft-state spectrum of MAXI J1820+070 is best characterized by an absorbed, relativistic thin accretion disk paired with a Comptonization tail and two distinct blackbody emission components with temperatures of $k T_{\text{bb}} \sim 0.2\text{ keV}$ and $\sim 0.5\text{ keV}$, respectively.

Our results indicate that the black hole in MAXI J1820+070 is rapidly rotating, with a spin constraint of $a > 0.8$. This finding is highly consistent with independent results from X-ray reflection spectroscopy and relativistic precession analysis \citep{Bhargava_2021, Draghis_2023}, but directly contradicts previous continuum-fitting estimates \citep{Zhao_2021, Guan_2021, Sai_2024}. The discrepancy stems from the omission of the two low-energy blackbody components in prior works. Specifically, the component with $k T_{\text{bb}} \approx 0.2\text{ keV}$ drastically improves the spectral fits below $0.5\text{ keV}$, while the second component at $k T_{\text{bb}} \approx 0.5\text{ keV}$ significantly resolves model-data discrepancies across the $1\text{--}5\text{ keV}$ range. Accounting for the existence of these two features ultimately drives the continuum-fitting model toward a considerably higher black hole spin estimation.  

Our analysis reveals a substantial drop in the inner disk temperature midway through the soft state, which manifests as an apparent black hole spin variation. This shift in the emission flux cannot be explained solely by changes in the accretion rate. Instead, the inferred spin evolution likely reflects a modest outward migration of the disk's inner boundary to a maximum radius of $3R_{\text{g}}$, particularly given that the color correction factor remains fairly stable at $f_{\text{col}} \approx 1.56$. This result is fully consistent with the three-dimensional GRMHD simulations of inner thin disk structures around rapidly rotating black holes presented by \citet{Hankla_2025}.

The dual blackbody soft excess detected below $10\text{ keV}$ is consistent with the presence of two warm, optically thick, and spatially extended coronal layers. These layers are characterized by radial extents of $\Delta R \sim 6R_{\text{g}}$ and $\Delta R \sim 35R_{\text{g}}$, with temperatures of $0.5\text{ keV}$ and $0.2\text{ keV}$, respectively. This interpretation is highly analogous to the warm Comptonization models widely invoked to explain the soft X-ray excess in active galactic nuclei. By applying the threshold condition for warm Comptonization (instead of Compton cooling), alongside the generalized spin-dependent temperature profile derived in \citet{Papavasileiou_2025}, we conclude that these warm coronal layers must reside above the disk at extended radii beyond $R \sim 10R_{\text{g}}$ and $R \sim 40R_{\text{g}}$, respectively.

\bibliography{references}

\begin{appendix}
\section{2D posterior probability contours}
\label{Append}

\begin{figure}[ht] 
\begin{tikzpicture}
\centering
  \node (img1)  {\includegraphics[width=\linewidth]{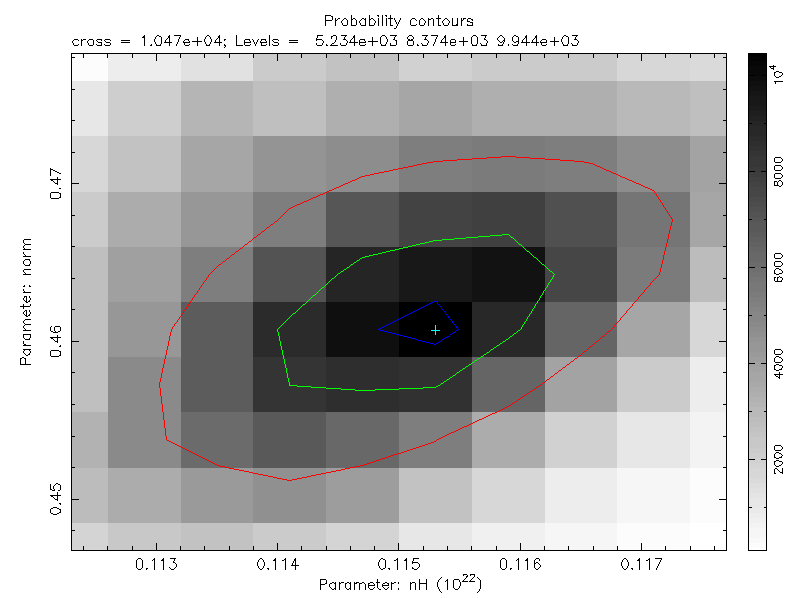}};
  \node[below= of img1, yshift=1cm] (img2)  {\includegraphics[width=\linewidth]{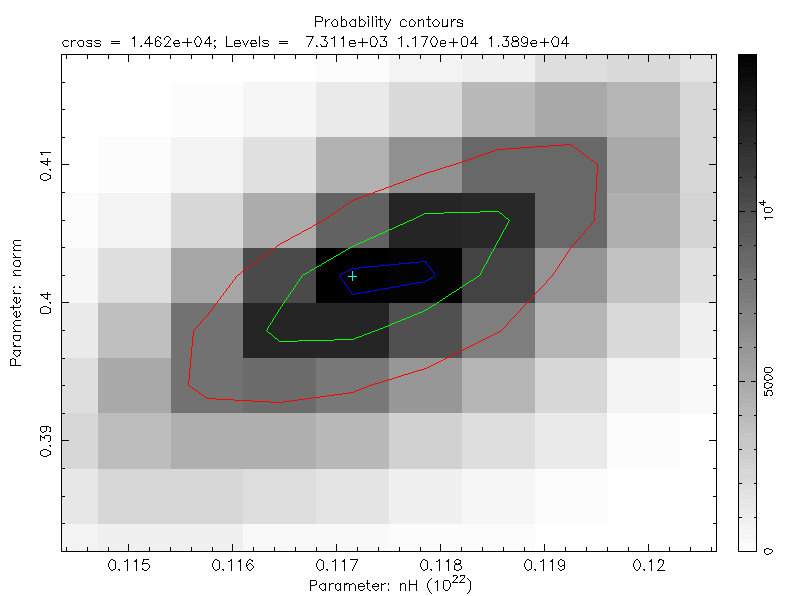}};
\end{tikzpicture}

\caption{\label{prob_contour} MCMC-derived 2D posterior probability contours for the hydrogen column density nH and the normalization of the $kT_{\text{bb}}=0.2\text{ keV}$ blackbody component, shown for the nu25–ni207 (top panel) and nu27–ni220 (bottom panel) observation epochs.} 
\end{figure}

\end{appendix}

\end{document}